\documentclass[aps,superscriptaddress,twocolumn,]{revtex4}
\usepackage{bm}
\usepackage{epsfig}
\usepackage{times}
\usepackage{bbm}
\usepackage{comment}
\usepackage{amssymb}
\usepackage{amsmath}
\usepackage{natbib}
\usepackage{color}
\usepackage[normalem]{ulem} 
\usepackage[latin1]{inputenc}
\usepackage[loose,nice]{units}       

\newcommand{\eq}[1]{Eq.~(\ref{#1})}
\begin{document}
\title{A non-relativistic Dirac equation: An application to photo ionization of highly charged hydrogen-like ions}
\author{Tor Kjellsson~Lindblom}
\affiliation{Deparment of Computer Science, OsloMet -- Oslo Metropolitan University, NO-0130 Oslo, Norway}
\author{Simen Br{\ae}ck}
\affiliation{Deparment of Computer Science, OsloMet -- Oslo Metropolitan University, NO-0130 Oslo, Norway}
\author{S{\o}lve Selst{\o}}
\email{solve.selsto@oslomet.no}
\affiliation{Deparment of Computer Science, OsloMet -- Oslo Metropolitan University, NO-0130 Oslo, Norway}

\begin{abstract}
We investigate the role of relativity in photo ionization of hydrogen-like ions by a laser pulse. For hydrogen, the wavelengths of the laser resides in the weakly ultra violet region. For higher nuclear charges, the laser parameters are scaled in a manner which renders the time-dependent Schr{\"o}dinger equation in the dipole approximation independent of nuclear charge. The ionization potentials of these highly charged ions are strongly modified by relativistic effects. In an earlier work, Ivanova et al., Phys. Rev. A {\bf 98}, 063402 (2018), it is demonstrated how this explains most of the relativistic correction to the ionization probability. Here we investigate to what extent remaining discrepancies can be attributed to relativistic effects stemming from the strong external field. To this end, we solve semi-relativistic formulations of both the Schr{\"o}dinger and the Dirac equations; the former accounts for increased inertia due to the external laser field, while the latter features a non-relativistic interaction term.
\end{abstract}

\maketitle

\section{Introduction}
\label{Introduction}

Motivated by experimental breakthroughs, we have seen an increased interest in the theoretical and computational study of relativistic dynamics for atoms and ions exposed to strong laser pulses \cite{DiPiazza2012}.
Such investigations tend to be quite demanding from a computational point of view as the ionization process is governed by the time-dependent Dirac equation, which typically is quite hard to solve. For this reason, many studies resort to models and approximations -- such as the  \textit{strong field approximation} and models of reduced dimensionality. However, also on the theoretical and computational side we have seen recent breakthroughs when it comes to fully relativistic descriptions, see, e.g., \cite{Braun99, FillionGourdeau2012, Ivanov15, FillionGourdeau16, Kjellsson17_MC, Kjellsson2017_PG, Ivanova18, KjellssonLindblom18, Tumakov2020, Vembe2024}.
Moreover, it has also been shown how exponential speedup in solving the Dirac equation may be achieved on a quantum computer \cite{FillionGourdeau17}.

The numerical solution of the time-dependent Dirac equation is subject to several issues which its non-relativistic counterpart, the time-dependent Schr{\"o}dinger equation, does not suffer from. For instance, many conventional numerical schemes for the time evolution is subject to very restrictive limitation on the numerical time step. This problem may, however, often be evaded by the use of so-called Magnus propagators \cite{Blanes2009}.

Another issue, which arises when using spectral methods, is the fact that the inclusion of the spatial dependence of external electromagnetic fields, such as laser pulses, is non-trivial \cite{Selsto2009, Kjellsson17_MC, Telnov2020}. However, it has been shown that this challenge to a large extent can be overcome by formulating the interaction in what is coined the {\it propagation gauge} \cite{Aldana2001, Forre16, Simonsen2016, Kjellsson2017_PG, Forre2020}. Another fruitful approach is that of expanding the external field in terms of a finite Fourier series~\cite{Vembe2024}. Also, within a non-relativistic framework, the notion of \textit{partial Fourier transforms} within a split operator scheme~\cite{Suster2023, Suster2024} seems a promising venue -- one that hopefully can be extended into the relativistic realm in the near future.

Other methods aim to incorporate relativistic corrections in a non-relativistic context by means of a semi-relativistic Schr{\"o}dinger equation. Substituting the electron's rest mass with a field-dressed, relativistic mass has proven to be a viable path in this direction~\cite{KjellssonLindblom18}. For strong fields, this approach has been used to describe photo ionization of atoms at photon energies in both the optical~\cite{Lindblom2020} and the X-ray~\cite{Forre2019, Forre2020} regime -- in addition to ultra-violet frequencies with general ellipticity~\cite{Lindblom2021}. Incidentally, it has also been shown that a similar mass-replacement may be introduced into a non-relativistic framework in order to account for the mass shift induced by internal energy-transitions for a moving atom~\cite{Sonnleitner18}.

In Ref.~\cite{Ivanova18} the photo ionization of various hydrogen-like ions by laser pulses was studied in a relativistic framework. For hydrogen, the photon energies involved corresponded to the weakly ultra violet region -- more specifically, with wavelengths ranging from 100 to 400~nm. For highly charged ions, strong relativistic effects where seen in the total ionization yield. It was convincingly argued that these corrections predominantly originated from relativistic corrections to the ionization potential of the ions as the nuclear charge $Z$ was increased. To this end, scaling relations was introduced which, to a large extent, were able to reproduce the ionization probabilities obtained for the non-relativistic hydrogen case, $Z=1$.
These relativistic calculations were performed within the so-called {\it dipole approximation}, in which the spatial dependence of the external laser field is neglected. While this approximation may be questioned in this regime, the argument that the increased ionization potential should lead to shifted, reduced ionization probability should still hold. However, the shift of the spectrum of the unperturbed ion does not explain the relativistic photo ionization yield in full. By direct solutions of the Schr{\"o}dinger equation and the Dirac equation -- within and without semi-relativistic approximations -- we set out to identify potential relativistic corrections induced by external field adding to relativistic effects stemming from the strong Coulomb attraction from the nucleus.

The paper is organized as follows: In Sec.~\ref{Theory} the theoretical framework and numerical implementations are outlined, and in Sec.~\ref{Results} the results are presented and discussed.
Conclusions are drawn in Sec.~\ref{Conclusion}. Atomic units are used where stated explicitly.

\section{Theory and implementation}
\label{Theory}

The time-dependent Dirac equation may, like its non-relativistic counterpart, the Schr{\"o}dinger equation, be written
\begin{equation}
\label{GenericEquation}
\mathrm{i} \hbar \frac{\mathrm{d}}{\mathrm{d} t} \Psi = H \Psi .
\end{equation}
The wave function $\Psi$ is a scalar function in the Schr{\"o}dinger equation, while the state is a four-component vector in the Dirac case. And the relativistic Hamiltonian $H^\mathrm{R}$ is a $4 \times 4$ matrix:
\begin{equation}
\label{TDDEham}
H^\mathrm{R} =  c \boldsymbol{\alpha} \cdot ({\bf p} + e{\bf A}) + V(r; Z) \, I_4
+ m c^2 \beta .
\end{equation}
Here,
\begin{equation}
\label{CoulombPot}
V(r;Z) = -  \frac{1}{4 \pi \varepsilon_0} \frac{Z}{r}
\end{equation}
is the Coulomb potential of the nucleus, which is assumed to be of infinite mass. For the $\alpha$-matrices, we have adopted the usual formulation in terms of Pauli matrices.
\begin{equation}
\label{AlphaDef}
\alpha_k =
\left( \begin{array}{cc} 0 & \sigma_k \\ \sigma_k & 0 \end{array} \right) \quad ,
\end{equation}
where $k$ is either $x$, $y$ or $z$,
and
\begin{equation}
\label{BetaDef}
\beta =
\left( \begin{array}{cc} I_2 & 0 \\ 0 & -I_2 \end{array} \right) \quad .
\end{equation}

The non-relativistic Schr{\"o}dinger Hamiltonian reads
\begin{subequations}
\label{SchrodHam}
\begin{align}
\label{SchrodHamFull}
H^\mathrm{NR} & = \frac{({\bf p} + e {\bf A} )^2}{2m} + V(r; Z) = \\
\label{SchrodHamPert}
 &
H^\mathrm{NR}_0 +\frac{e}{m} {\bf A} \cdot {\bf p} + \frac{e^2}{2m} {\bf A}^2,
\end{align}
\end{subequations}
where $H^\mathrm{NR}_0$ is the time-independent part of the non-relativistic Schr{\"o}dinger Hamiltonian.

The external laser pulse enters via the vector potential ${\bf A}$. In our numerical studies, we impose the dipole approximation, i.e., we will take ${\bf A}$ to be purely time-dependent. Consequently, the electric field will be homogeneous and the magnetic field will be absent. While this approximation may be challenged in the regime in which we aim to apply it, this does not prevent us from illustrating how the alternative, semi-relativistic formulations of the interaction enables us to distinguish between structural and dynamical relativistic effects.

\subsection{Relativistic mass in the Schr{\"o}dinger equation}

In Ref.~\cite{KjellssonLindblom18} it explained how the leading relativistic corrections induced by the external field ${\bf A}$ may be accounted for within the Schr{\"o}dinger equation by replacing the rest mass $m$ by the field-dressed effective mass. This is done within the relativistic version of the \textit{propagation gauge}~\cite{Kjellsson2017_PG} with the spatial dependence of the vector potential retained.

Within the dipole approximation it simplifies a bit, and the expressions for the field-dressed mass reads
\begin{equation}
\label{RelativisticMass}
\mu(t) = m \sqrt{1 + \left( \frac{{e {\bf A}}}{mc} \right)^2} .
\end{equation}
If we neglect spin and relativistic corrections to the Coulomb potential, the mass-adjusted semi-relativistic Hamiltonian becomes a rather direct replacement of $m$ by $\mu(t)$ in Eq.~(\ref{SchrodHamFull}) in the dipole approximation:
\begin{subequations}
\label{SemiRelSchrod}
\begin{align}
\label{SemiRelSchrodFull}
H^\mathrm{SR}_\mathrm{S} & =
\frac{\mathbf{p}^2}{2 \mu(t)} + \frac{e}{\mu(t)} \mathbf{A} \cdot \mathbf{p}  +
\frac{e^2}{2m} \mathbf{A}^2 + V(r; Z) =\\
\label{SemiRelSchrodPert}
 &
H^\mathrm{NR}_0 +\left(\frac{1}{\mu(t)}-\frac{1}{m} \right)  \frac{ {\bf p}^2 }{2}
+\frac{e}{\mu(t)} {\bf A} \cdot {\bf p} + \frac{e^2}{2m} {\bf A}^2 .
\end{align}
\end{subequations}
Note that the correction to the kinetic energy term in effect introduces an additional interaction term as compared to Eq.~(\ref{SchrodHamPert}).

The notion of a field-dressed, effecive mass enables us to account, to leading order, for strong-field relativistic effects induced by the external field accelerating the electron towards a significant fraction of the speed of light. Correspondingly, a comparison of the solutions of the Schr{\"o}dinger equation with the Hamiltonians of Eqs.~(\ref{SchrodHam}) and (\ref{SemiRelSchrod}) should reveal such effects.

\subsection{Non-relativistic interaction in the Dirac equation}
\label{Interactions}

If we separate the Hamiltonian of \eq{TDDEham} in a time-dependent and a time-independent part, it reads
\begin{equation}
\label{Hseparate}
H^\mathrm{R} = H_0^\mathrm{R} + c \boldsymbol{\alpha} \cdot {\bf A} \quad .
\end{equation}
In Ref.~\cite{Selsto2009} an alternative, non-relativisic form of the interaction was presented.
For completeness, we will here briefly revise the derivation of this formulation, which is attributable to prof. Eva Lindroth.

By introducing the Foldy-Wouthuysen-like unitary transformation~\cite{Foldy50}
\begin{equation}
\label{EvaGaugeTransform}
\Psi' = U \Psi \quad \text{with} \quad U=\exp \left[ \frac{e}{2mc} \beta \boldsymbol{\alpha} \cdot {\bf A} \right] ,
\end{equation}
and formulating the Dirac equation, \eq{GenericEquation}, in terms of $\Psi'$ rather than $\Psi$, we arrive at the effective Hamiltonian
\begin{equation}
\label{UntruncatedEvaHam}
H^\mathrm{SR}_\mathrm{D} = U H^\mathrm{R} U^\dagger + \mathrm{i} \hbar \, \frac{\partial U}{\partial t} U^\dagger  \quad .
\end{equation}
If we expand $U$ up to second order, we will find that $H^\mathrm{SR}_\mathrm{D}$ may be written
\begin{align}
\label{ExpandHprime}
 H^\mathrm{SR}_\mathrm{D}
 & =
 c \boldsymbol{\alpha} \cdot {\bf p} +
 \frac{e}{m} {\bf A} \cdot {\bf p} \, \beta + \frac{e^2}{2m} A^2 \, \beta +
 \frac{e \hbar}{2m} \boldsymbol{\sigma} \cdot {\bf B} \, \beta
 \\ \nonumber &
 + V(r; Z) + \mathcal{O}({c^{-2}}) \quad ,
\end{align}
which holds beyond the dipole approximation as well.
In arriving at Eq.~(\ref{ExpandHprime}), we impose the Couloumb gauge restriction: $\nabla \cdot {\bf A} = 0$. Also, the identity
\begin{equation}
(\boldsymbol{\alpha} \cdot {\bf a}) \, (\boldsymbol{\alpha} \cdot {\bf a}) =
{\bf a} \cdot {\bf b} + i \boldsymbol{\sigma} \cdot ( {\bf a} \times {\bf b})
\end{equation}
is useful in deriving this semi-classical form of the interaction.

Within the dipole approximation, the magnetic field ${\bf B} = \nabla \times {\bf A}$ vanishes, and we arrive at the following semi-relativistic formulation:
\begin{subequations}
\label{SemiRelDirac}
    \begin{align}
        \label{SemiRelDiracFull}
        H^\mathrm{SR}_\mathrm{D} & = c \boldsymbol{\alpha} \cdot {\bf p} +
        \left(
        \frac{e}{m} {\bf A} \cdot {\bf p} + \frac{e^2}{2m} {\bf A}^2 \right) \beta
        + V(r; Z) \\
        \label{SemiRelDiracPert}
        & = H^\mathrm{R}_\mathrm{0} + \left(
        \frac{e}{m} {\bf A} \cdot {\bf p} + \frac{e^2}{2m} {\bf A}^2 \right) \beta
    \end{align}
\end{subequations}

Eq.~(\ref{SemiRelDirac}) is semi-classical in the sense that it features the interaction terms of the non-relativistic Schr{\"o}dinger Hamiltonian within the minimal coupling formulation.
Apart from the $\beta$-matrix, the interaction terms of \eq{SemiRelDiracPert} coincides with those of \eq{SchrodHamPert}.

The crucial difference between \eq{SemiRelDiracPert} and \eq{SchrodHamPert} is their time-independent parts, $H_0$. In the former case, this part is fully relativistic and, thus, entirely capable of resolving the relativistic structure of the system.
Relativistic effects induced by the external laser field, on the other hand, are \textit{not} resolved by this formulation as it features a non-relativistic formulation of the interaction. Correspondingly, by comparing the predictions within the two different formulations of the Dirac Hamiltonian, we should be able to distinguish between structural and dynamical relativistic effects -- between relativistic corrections induced, respectively, by the Coulomb interaction alone on one hand and the external laser field -- or the interaction between the two -- on the other.


\subsection{Scaling relations}
\begin{figure}
    \centering
    \includegraphics[width=1\linewidth]{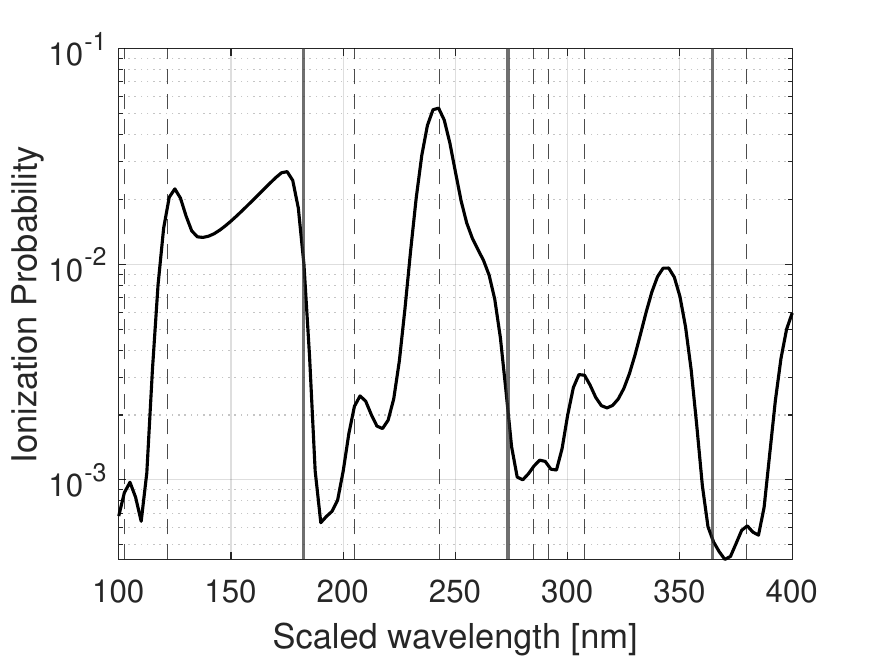}
    \caption{The ionization probability for a hydrogen-like ion as a function of the scaled wavelength $\tilde{\lambda}$. The scaled maximum field strength is $\tilde{E}_0 =0.03$~a.u., which corresponds to an intensity of $3.16\cdot10^{13}$~W/cm$^2$ for hydrogen. The thick, gray vertical lines corresponds, from left to right, to the onset of the two, three and four photon ionization thresholds. The dashed lines corresponds to wavelengths relevant for REMPI processes from the ground state. With the notation $(n_\mathrm{int}, n_\gamma)$ indicating an $n_\gamma$-photon transition via the state with principal quantum number $n_\mathrm{int}$, these correspond, from left to right, to $(3, 1)$, $(2, 1)$, $(3, 2)$, $(2, 2)$, $(5, 3)$, $(4, 3)$, $(3, 3)$ and $(5, 4)$.}
    \label{Fig_REMPI}
\end{figure}

It has been shown that by introducing the appropriate scaling of the variables and laser parameters, the time-dependent Schr{\"o}dinger equation for the laser-matter interaction becomes independent of the nuclear charge within the dipole approximation \cite{Lambropoulos1999}.
We will identify these scaled variables and parameters by a tilde,
\begin{subequations}
\label{Scaling}
\begin{align}
\label{ScalingR}
&\tilde{r}=Z r \\
\label{ScalingT}
&\tilde{t}=Z^2 t \\
\label{ScalingE0}
&\tilde{E}_0 = E_0/Z^3 \quad \text{and} \\
\label{ScalingOmega}
& \tilde{\omega} = \omega/Z^2 \quad .
\end{align}
\end{subequations}
As an illustration, we demonstrate the ionization probability as a function of the scaled wavelength $\tilde{\lambda}= \lambda / Z^2$ in Fig.~\ref{Fig_REMPI}. For a hydrogen atom, the wavelengths falls into the moderately ultra-violet region. To the extent that the dipole approximation and the non-relativistic treatment is valid, this plot is independent of $Z$. Here, the scaled peak intensity is $\tilde{I} = 3.16\cdot 10^{13}$~a.u. According to Eq.~(\ref{ScalingE0}), the actual peak intensity $I = Z^6 \tilde{I}$.

As explained in Ref.~\cite{Ivanova18}, the strongly non-monotonic behaviour is due to the onset of the two, three and three-photon thresholds, accompanied by pronounced peaks whenever
photon absorption happens to be in resonance with intermediate excited bound states (\textit{resonantly enhanced multi photon ionization} - REMPI). In Fig.~\ref{Fig_REMPI} we have indicated both thresholds and REMPI-energies by vertical lines.

The $Z$-independence does not survive any transition beyond the dipole approximation nor to any relativistic description. Thus,
any deviation from the  scaling law of Eqs.~(\ref{Scaling}) is indicative of the breakdown of the dipole approximation and/or the non-relativistic approach -- the latter in the present case.

It is worth mentioning that in Ref.~\cite{Ivanova18} another set of approximate scaling relations which also accounted for the relativistic shift in ionization potential was imposed. These scaling relations were quite successful in explaining the photo ionization's dependence of laser wavelength within the dipole approximation.
In this context, however, we will resort to the scaling of Eqs.~(\ref{Scaling}).

\subsection{Implementation}

A spectral method is applied for resolving the time evolution dictated by Eq.~(\ref{GenericEquation}). The wave functions are expressed in terms of basis functions consisting of the eigenstates of the time-independent part of the Hamiltonians, $H_0^\mathrm{R/NR}$, by means of expansions in B-splines~\cite{deboor} and spherical harmonics. Specifically, for the time-independent Dirac Hamiltonian, the basis function have the form
\begin{subequations}
\label{RelExpansion}
\begin{align}
\label{RelExpansionPsi}
& \psi^\mathrm{R}_{n,j,m_j,\kappa}  =  \frac{1}{r} \left(\begin{array}{cc} P_{n,\kappa}(r) X_{j,m_j,\kappa}(\Omega) \\
i Q_{n,\kappa}(r) X_{j,m_j,-\kappa}(\Omega) \end{array}
 \right)
\\
\label{RelExpansionP}
&
P_{n,j,\kappa}(r) = \sum_i a_{n,\kappa}^i B_i^{k_1}(r),
\\
&
\label{RelExpansionQ}
Q_{n,j,\kappa}(r) = \sum_i b_{n,\kappa}^i B_i^{k_2}(r) ,
\\
&
\label{RelExpansionX}
X_{j,m_j,\kappa}(\Omega) =
\sum_{m_\ell, m_s} \langle \ell, m_\ell, 1/2, m_s | j, m_j\rangle \chi_{m_s} Y_{\ell,m_\ell}(\Omega),
\\
&
\label{RelExpansionKappa}
\kappa=
\left\{
\begin{array}{ll}
\ell, & j=\ell-1/2 \\ -(\ell+1), & j=\ell+1/2
\end{array}
\right. ,
\end{align}
\end{subequations}
where $B_i^k(r)$ is a B-spline of order $k$ and $\chi_{m_s}$ is an eigenspinor.
We impose Dirichlet boundary conditions at $r_\mathrm{max} = 250/Z$ a.u..

The non-relativistic basis functions are simpler:
\begin{equation}
\label{NonRelExpansion}
\psi^\mathrm{NR}_{n, \ell, m_\ell} =
\frac{1}{r} P_{n, \ell}(r) Y_{\ell, m_\ell}(\Omega) \quad ,
\end{equation}
where the radial part $P_{n, \ell}(r)$ is expanded analogous to Eq.~(\ref{RelExpansionP}).

For linearly polarized fields pointing along the $z$-axis described within the dipole approximation, we need only include angular components within the same $m$ quantum number, be it $m_j$ or $m_\ell$, as the initial ground state. We include all angular momenta $\ell$ up to $\ell_\mathrm{max}=20$, as well as all states with energies below $mc^2 + 100Z^2$ a.u. for the positive part of the spectrum and above $-mc^2 - 100Z^2$ for the negative part; prior works have shown the negative part of the spectrum cannot be neglected in general, see, e.g., \cite{Selsto2009, Vanne2012}.
The numerical solution of the time-independent Dirac equation is typically plagued with the emergence of so-called \textit{spurious states}. For the present method, Ref.~\cite{Fischer2009} has presented detailed evidence that the choice $k_1=k_2\pm 1$ for the order or the B-splines used to expand $P$ and $Q$ in Eqs.~(\ref{RelExpansionP}) and (\ref{RelExpansionP}), respectively, successfully resolves this issue. Here, we have chosen the values $k_1=7$ and $k_2=8$, which, in our experience, provide good accuracy for these kind of calculations \cite{Kjellsson17_MC,Kjellsson2017_PG}.

For a Coulomb potential strong enough to induce relativistic effects, a uniform knot sequence is highly impractical in solving the time-independent Dirac equation. Instead we have applied the same type of sequence as in Ref.~\cite{Vanne2012}: the knot points are first distributed in a geometric sequence which switches to a linear distribution at a certain point. We found that $N_{\mathrm{knot}}=759$ with geometric factor $1.05$ and switching point at the 280th knot point gives well-converged good numerical spectra for the relativistic calculations in this paper.

For the solution of the time-dependent Dirac equations, we need to calculate coupling elements of form
\begin{equation}
\label{GenericCoupl}
\langle \psi^\mathrm{R}_{n, j, n_j, \kappa} | H_I | \psi^\mathrm{R}_{n', j', m_j', \kappa'} \rangle  ,
\end{equation}
where $H_I$ would correspond to $\alpha_z$ in the case of Eq.~(\ref{Hseparate}) and $p_z \beta$ and $\beta$, respectively, in the case of Eq.~(\ref{SemiRelDirac}). Each of these coupling elements is to be multiplied with their appropriate time-dependent factor. The corresponding non-relativistic matrix elements,
\begin{equation}
\label{GenericCouplNonRel}
\langle \psi^\mathrm{NR}_{n, \ell, m_\ell} | H_I | \psi^\mathrm{NR}_{n, \ell', m_\ell'} \rangle  ,
\end{equation}
has $H_I \sim p_z$ for the Hamiltonian of Eq.~(\ref{SchrodHamPert}) and ${\bf p}^2$ and $p_z$, respectively, for Eq.~(\ref{SemiRelSchrodPert}). In both cases, the term proportional to ${\bf A}^2$ may safely be ignored since we work within the dipole approximation.
The radial integrals involved in calculating the coupling elements of Eqs.~(\ref{GenericCoupl}) and (\ref{GenericCouplNonRel}) are evaluated by a Gauss-Legendre quadrature.

Next, the time evolution is carried out by propagating the state vector using a Magnus type propagator \cite{Blanes2009}:
\begin{equation}
\label{MagnusProp}
\Psi(t+\tau) \approx \exp(-i/\hbar \, H(t+\tau/2) \tau )\Psi(t) + \mathcal{O}(\tau^3).
\end{equation}
This evades the extreme restriction that the numerical time step must be significantly lower than the inverse of the electron's mass energy, which several other propagation schemes suffer from when solving the time-dependent Dirac equation.
On the other hand, repeatedly exponentiating large matrices is not tractable either. To this end, we have approximated the action of the Magnus propagator in Eq.~(\ref{MagnusProp}) by projecting it into the time-dependent Krylov subspace of dimension $k$,
\begin{align}
\label{KrylovApproximation}
& \exp(-i/\hbar \, H(t+\tau/2) \tau ) \Psi(t)  \approx
\\ \nonumber &
V \exp(-i/\hbar \, H_\mathcal{K}(t+\tau/2) \tau ) V^{\dagger}  \Psi(t) ,\end{align}
where the orthonormal Krylov basis consists of the columns of the projection matrix $V$. These, in turn, are constructed from the set obtained from the repeated series of matrix-vector multiplications
\begin{equation}
\label{ConstructKrylovBasis}
 \left[H(t+\tau/2)\right]^m \Psi(t), \quad m = 0, \ldots, k-1
\end{equation}
by the Arnoldi procedure with reorthogonalization.
The Krylov representation of the Hamiltonian matrix, $H_\mathcal{K}$, is a $k \times k$ matrix, where $k$ is considerably lower than the full dimensionality of the numerical problem. Thus,
the most time consuming part is not the exponentiation but rather the repeated series of matrix-vector multiplications, Eq.~(\ref{ConstructKrylovBasis}).
These iterations are performed until the estimated error of the propagated wave is below some specified limit \cite{Saad1992}.

For all cases presented here we have checked for convergence in all numerical parameters. It is found that 450-600 numerical time steps per optical cycle suffices for the Krylov propagators to converge within $k=15$ iterations for the non-relativistic calculations and 30 iterations for the relativistic ones per time step.
The factor two in the relativistic Krylov subspace size is directly related to the stiffness induced by the mass energy term, i.e., the last term in the Hamiltonian of Eq.~(\ref{TDDEham}).
For the Dirac equation
our predictions were compared with the corresponding results of Ref.~\cite{Ivanova18} and found to be in agreement.

\section{Results and Discussion}
\label{Results}

As mentioned, we take our laser field to be linearly polarized along the $z$-axis. Moreover, we use a model for the homogeneous laser pulse of this shape:
\begin{equation}
\label{LaserPulse}
A(t) = \frac{E_0}{\omega} \, \sin^2\left( \frac{\pi}{T} \, t\right) \sin(\omega t) , \quad t \in [0, T] \quad .
\end{equation}
The pulse duration corresponds to 20 optical cycles.
Although the dipole approximation may be challenged for the systems at hand, our calculations should still enable us to distinguish relativistic effects induced by the Coulomb field alone from those stemming from the external laser field.

\begin{figure}
    \centering
    \begin{tabular}{c}
    \includegraphics[width=1\linewidth]{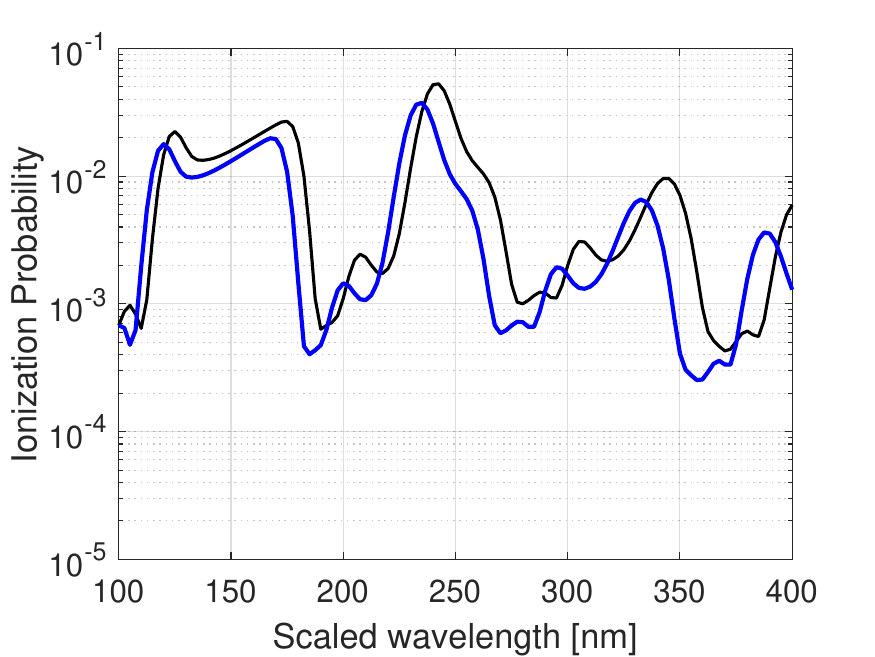} \\
    \includegraphics[width=1\linewidth]{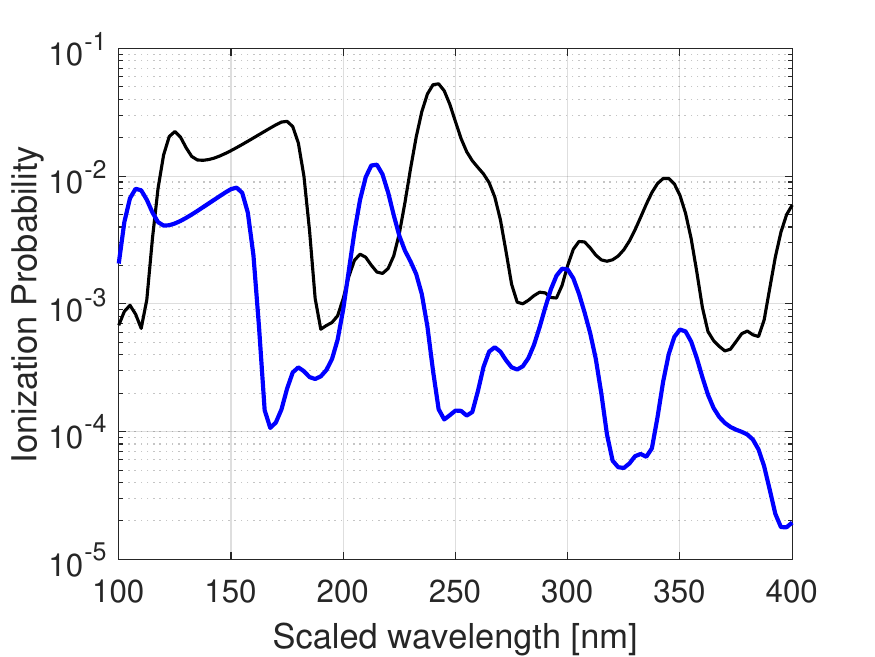}
    \end{tabular}
    \caption{The ionization probability for a hydrogen-like ion with nuclear charge $Z=50$ (upper) and $Z=92$ (lower). The scaled field parameters are the same as in Figs.~\ref{Fig_REMPI}. The black curve is the $Z$-independent prediction from the Schr{\"o}dinger equation within the dipole approximation while the blue curve is obtained solving the full Dirac equation within the dipole approximation.}
    \label{Fig_TDDE}
\end{figure}

For the hydrogen case, no noticeable difference between non-relativistic, semi-relativistic or fully relativistic calculations are seen. For higher nuclear charges, however, we see strong deviations between the predictions of the Schr{\"o}dinger equation and the Dirac equation. Figure~\ref{Fig_TDDE} shows the relativistic ionization probability for the same case as in Fig.~\ref{Fig_REMPI} with $Z=50$ and $Z=92$. Although the laser intensity is somewhat higher, what we see is in perfect alignment with the findings of~\cite{Ivanova18}: the ionization probability is shifted towards shorter wavelengths and there is an over all down-shift in ionization yield. The former was convincingly explained in terms of increased ionization potential. To some extent, this relativistic shift can also, via the introduction of an effective nuclear charge, explain the decrease in over all yield. But not fully.

\begin{figure}
    \centering
    \begin{tabular}{c}
    \includegraphics[width=1\linewidth]{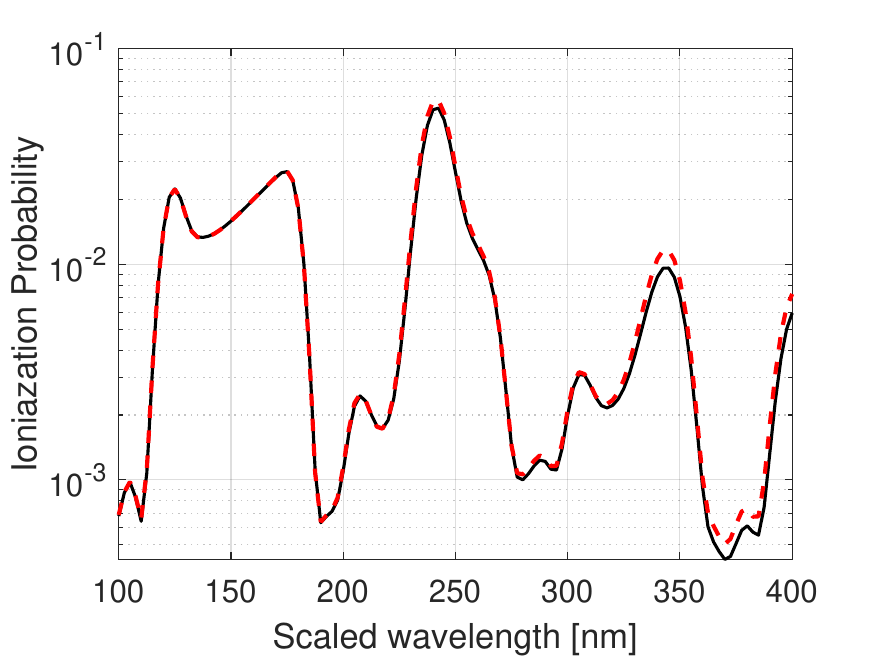} \\
    \includegraphics[width=1\linewidth]{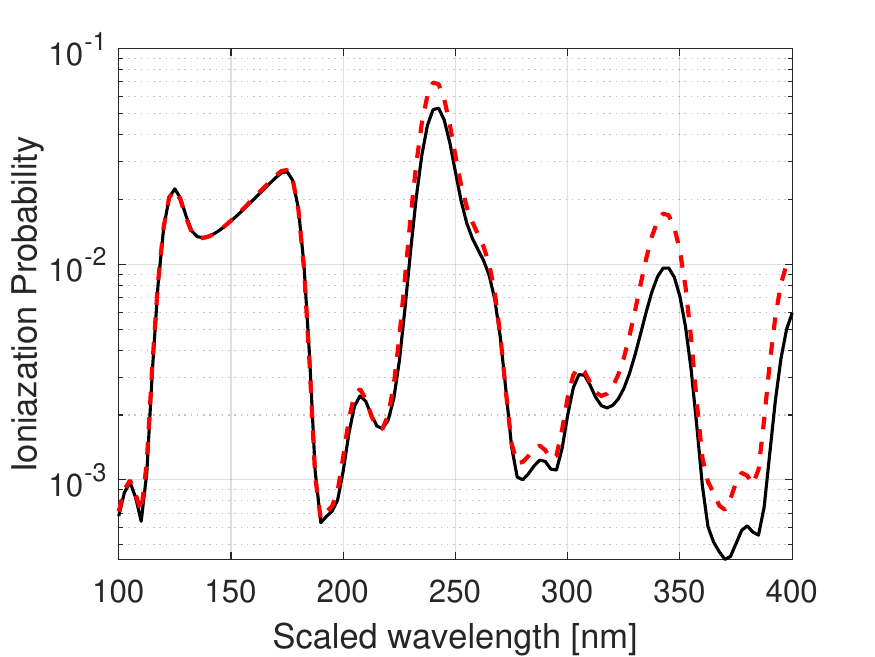}
    \end{tabular}
    \caption{The ionization probability for a hydrogen-like ion with nuclear charge $Z=50$ (upper) and $Z=92$ (lower). The scaled field parameters are the same as in Figs.~\ref{Fig_REMPI} and \ref{Fig_TDDE}. As before, the black curve is result from the Schr{\"o}dinger equation while the red, dashed curve is obtained using the semi-relativistic Schr{\"o}dinger Hamiltonian of Eq.~(\ref{SemiRelSchrod}).}
    \label{Fig_SR-TDSE}
\end{figure}

To investigate this further, we have solved the semi-relativistic Schr{\"o}dinger equation, i.e., we have solved Eq.~(\ref{GenericEquation}) using the Hamiltonian of Eq.~(\ref{SemiRelSchrod}). While this will not be able to explain the shift towards shorter wavelengths, one may hope that the decrease in ionization yield can be understood in terms of increased inertia induced by the laser pulse. The results are shown in Fig.~\ref{Fig_SR-TDSE} for the same cases as in Fig.~\ref{Fig_TDDE}.

As it turns out, introducing the relativistic mass has actually led to an \textit{increase} of the ionization yield. This may appear surprising since the replacement $m \rightarrow \mu(t)$, in effect, reduces the strength of the laser field in the dipole interaction term:
\begin{equation}
\label{Aeff}
\frac{e}{m} {\bf A} \cdot {\bf p} \rightarrow
\frac{e}{m} {\bf A}_\mathrm{eff} \cdot {\bf p} \quad \text{with} \quad
{\bf A}_\mathrm{eff}(t) = \frac{m}{\mu(t)} {\bf A}(t) .
\end{equation}
However, as also the kinetic energy term is affected by the field dressed mass, a second ionization mechanism is introduced, see Eq.~(\ref{SemiRelSchrodPert}). Apparently, this mechanism compensates any decrease that comes about via the reduced effective field strength

\begin{figure}
    \centering
    \begin{tabular}{c}
    \includegraphics[width=1\linewidth]{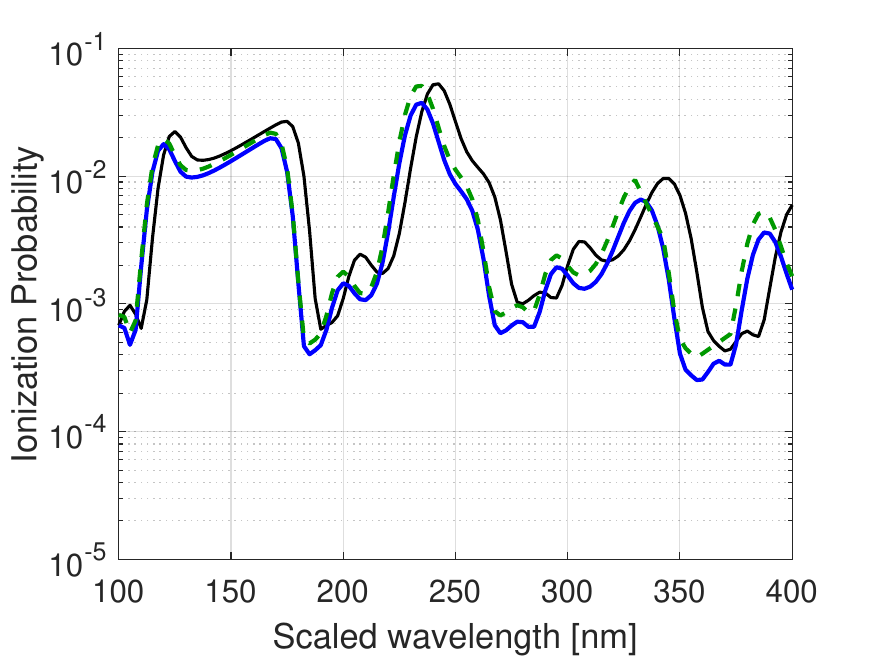} \\
    \includegraphics[width=1\linewidth]{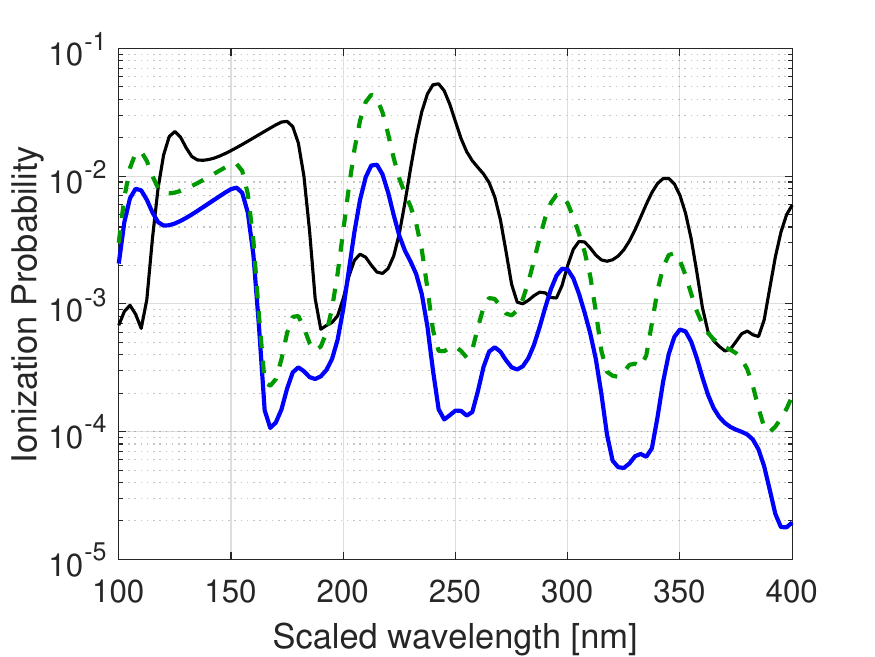}
    \end{tabular}
    \caption{This figure shows the same as Fig.~\ref{Fig_TDDE} -- with the additional inclusion of results obtained with the semi-relativistic Dirac Hamiltonian in Eq.~(\ref{SemiRelDirac}) (green, dashed curves).}
    \label{Fig_SR-TDDE}
\end{figure}

Next, we implement the other semi-relativistic approach -- the one in which the time-independent part remains fully relativistic, Eq.~(\ref{SemiRelDirac}). The resulting ionization probability is shown in Fig.~\ref{Fig_SR-TDDE}. With a fully relativistic spectrum for the unperturbed system, it should come as no surprise that the shift towards lower wavelengths is reproduced. More interestingly, the Dirac Hamiltonian with non-relativistic interaction overestimates the ionization yield.

Despite the conclusion in regard to Fig.~\ref{Fig_SR-TDSE}, we will still argue that this can be understood in terms of increased inertia. However, the effective, higher mass, $\mu$ in Eq.~(\ref{Aeff}), is not attributable to the external field. Since the internal Coulomb field drives the electron towards relativistic speeds, also this contributes to an increased inertia which, in effect, renders the external field experienced by the electron weaker. The proper Dirac Hamiltonian, Eq.~(\ref{TDDEham}), accounts for this, the semi-relativistic formuation, Eq.~(\ref{SemiRelDirac}), does not.

\section{Conclusions}
\label{Conclusion}

We studied the role of relativistic corrections in photo ionization of hydrogen-like ions by comparatively strong laser fields. For hydrogen, the wavelengths ranged from the weakly ultra violet region up to the onset of the optical region, while, in order to facilitate comparison, the field parameters were scaled appropriately with the nuclear charge for ions.

We took advantage of a semi-relativistic approximation of the Dirac Hamiltonian to study the nature of the relativistic corrections -- an approximation which combined a fully relativistic spectrum with non-relativistic interaction terms. While it was well known that the increased ionization potential of highly charged ions accounted for most of the relativistic shifts, we also identified how the increased inertia of the bound electron leads to an effective stabilization; since the Coulomb field of highly charged nuclei accelerates the electron towards relativistic speeds which, in effect, renders the external electric field of the laser weaker than would be the case for a non-relativistic electron.

\section*{Acknowledgements}

We would like to thank the authors of Ref.~\cite{Ivanova18} for providing us with data for benchmarking our implementation. We also thank Sigma2 -- the National Infrastructure for High Performance Computing and Data storage in Norway -- for providing the resources necessary for implementing our calculations  (Project No. NN9417K).
Also, fruitful discussions with prof. Eva Lindroth and prof.  Morten F{\o}rre are gratefully acknowledged.



\begin{thebibliography}{31}
\expandafter\ifx\csname natexlab\endcsname\relax\def\natexlab#1{#1}\fi
\expandafter\ifx\csname bibnamefont\endcsname\relax
  \def\bibnamefont#1{#1}\fi
\expandafter\ifx\csname bibfnamefont\endcsname\relax
  \def\bibfnamefont#1{#1}\fi
\expandafter\ifx\csname citenamefont\endcsname\relax
  \def\citenamefont#1{#1}\fi
\expandafter\ifx\csname url\endcsname\relax
  \def\url#1{\texttt{#1}}\fi
\expandafter\ifx\csname urlprefix\endcsname\relax\def\urlprefix{URL }\fi
\providecommand{\bibinfo}[2]{#2}
\providecommand{\eprint}[2][]{\url{#2}}

\bibitem[{\citenamefont{Di~Piazza et~al.}(2012)\citenamefont{Di~Piazza,
  M\"uller, Hatsagortsyan, and Keitel}}]{DiPiazza2012}
\bibinfo{author}{\bibfnamefont{A.}~\bibnamefont{Di~Piazza}},
  \bibinfo{author}{\bibfnamefont{C.}~\bibnamefont{M\"uller}},
  \bibinfo{author}{\bibfnamefont{K.~Z.} \bibnamefont{Hatsagortsyan}},
  \bibnamefont{and} \bibinfo{author}{\bibfnamefont{C.~H.}
  \bibnamefont{Keitel}}, \bibinfo{journal}{Rev. Mod. Phys.}
  \textbf{\bibinfo{volume}{84}}, \bibinfo{pages}{1177} (\bibinfo{year}{2012}).

\bibitem[{\citenamefont{Braun et~al.}(1999)\citenamefont{Braun, Su, and
  Grobe}}]{Braun99}
\bibinfo{author}{\bibfnamefont{J.~W.} \bibnamefont{Braun}},
  \bibinfo{author}{\bibfnamefont{Q.}~\bibnamefont{Su}}, \bibnamefont{and}
  \bibinfo{author}{\bibfnamefont{R.}~\bibnamefont{Grobe}},
  \bibinfo{journal}{Phys. Rev. A} \textbf{\bibinfo{volume}{59}},
  \bibinfo{pages}{604} (\bibinfo{year}{1999}).

\bibitem[{\citenamefont{Fillion-Gourdeau
  et~al.}(2012)\citenamefont{Fillion-Gourdeau, Lorin, and
  Bandrauk}}]{FillionGourdeau2012}
\bibinfo{author}{\bibfnamefont{F.}~\bibnamefont{Fillion-Gourdeau}},
  \bibinfo{author}{\bibfnamefont{E.}~\bibnamefont{Lorin}}, \bibnamefont{and}
  \bibinfo{author}{\bibfnamefont{A.~D.} \bibnamefont{Bandrauk}},
  \bibinfo{journal}{Comput. Phys. Commun.} \textbf{\bibinfo{volume}{183}},
  \bibinfo{pages}{1403 } (\bibinfo{year}{2012}).

\bibitem[{\citenamefont{Ivanov}(2015)}]{Ivanov15}
\bibinfo{author}{\bibfnamefont{I.~A.} \bibnamefont{Ivanov}},
  \bibinfo{journal}{Phys. Rev. A} \textbf{\bibinfo{volume}{91}},
  \bibinfo{pages}{043410} (\bibinfo{year}{2015}).

\bibitem[{\citenamefont{Fillion-Gourdeau
  et~al.}(2016)\citenamefont{Fillion-Gourdeau, Lorin, and
  Bandrauk}}]{FillionGourdeau16}
\bibinfo{author}{\bibfnamefont{F.}~\bibnamefont{Fillion-Gourdeau}},
  \bibinfo{author}{\bibfnamefont{E.}~\bibnamefont{Lorin}}, \bibnamefont{and}
  \bibinfo{author}{\bibfnamefont{A.}~\bibnamefont{Bandrauk}},
  \bibinfo{journal}{J. Comput. Phys.} \textbf{\bibinfo{volume}{307}},
  \bibinfo{pages}{122 } (\bibinfo{year}{2016}).

\bibitem[{\citenamefont{Kjellsson
  et~al.}(2017{\natexlab{a}})\citenamefont{Kjellsson, Selst\o{}, and
  Lindroth}}]{Kjellsson17_MC}
\bibinfo{author}{\bibfnamefont{T.}~\bibnamefont{Kjellsson}},
  \bibinfo{author}{\bibfnamefont{S.}~\bibnamefont{Selst\o{}}},
  \bibnamefont{and} \bibinfo{author}{\bibfnamefont{E.}~\bibnamefont{Lindroth}},
  \bibinfo{journal}{Phys. Rev. A} \textbf{\bibinfo{volume}{95}},
  \bibinfo{pages}{043403} (\bibinfo{year}{2017}{\natexlab{a}}).

\bibitem[{\citenamefont{Kjellsson
  et~al.}(2017{\natexlab{b}})\citenamefont{Kjellsson, F\o{}rre, Simonsen,
  Selst\o{}, and Lindroth}}]{Kjellsson2017_PG}
\bibinfo{author}{\bibfnamefont{T.}~\bibnamefont{Kjellsson}},
  \bibinfo{author}{\bibfnamefont{M.}~\bibnamefont{F\o{}rre}},
  \bibinfo{author}{\bibfnamefont{A.~S.} \bibnamefont{Simonsen}},
  \bibinfo{author}{\bibfnamefont{S.}~\bibnamefont{Selst\o{}}},
  \bibnamefont{and} \bibinfo{author}{\bibfnamefont{E.}~\bibnamefont{Lindroth}},
  \bibinfo{journal}{Phys. Rev. A} \textbf{\bibinfo{volume}{96}},
  \bibinfo{pages}{023426} (\bibinfo{year}{2017}{\natexlab{b}}).

\bibitem[{\citenamefont{Ivanova et~al.}(2018)\citenamefont{Ivanova, Shabaev,
  Telnov, and Saenz}}]{Ivanova18}
\bibinfo{author}{\bibfnamefont{I.~V.} \bibnamefont{Ivanova}},
  \bibinfo{author}{\bibfnamefont{V.~M.} \bibnamefont{Shabaev}},
  \bibinfo{author}{\bibfnamefont{D.~A.} \bibnamefont{Telnov}},
  \bibnamefont{and} \bibinfo{author}{\bibfnamefont{A.}~\bibnamefont{Saenz}},
  \bibinfo{journal}{Phys. Rev. A} \textbf{\bibinfo{volume}{98}},
  \bibinfo{pages}{063402} (\bibinfo{year}{2018}).

\bibitem[{\citenamefont{Kjellsson~Lindblom
  et~al.}(2018)\citenamefont{Kjellsson~Lindblom, F\o{}rre, Lindroth, and
  Selst\o{}}}]{KjellssonLindblom18}
\bibinfo{author}{\bibfnamefont{T.}~\bibnamefont{Kjellsson~Lindblom}},
  \bibinfo{author}{\bibfnamefont{M.}~\bibnamefont{F\o{}rre}},
  \bibinfo{author}{\bibfnamefont{E.}~\bibnamefont{Lindroth}}, \bibnamefont{and}
  \bibinfo{author}{\bibfnamefont{S.}~\bibnamefont{Selst\o{}}},
  \bibinfo{journal}{Phys. Rev. Lett.} \textbf{\bibinfo{volume}{121}},
  \bibinfo{pages}{253202} (\bibinfo{year}{2018}).

\bibitem[{\citenamefont{Tumakov et~al.}(2020)\citenamefont{Tumakov, Telnov, and
  Plunien}}]{Tumakov2020}
\bibinfo{author}{\bibfnamefont{D.~A.} \bibnamefont{Tumakov}},
  \bibinfo{author}{\bibfnamefont{D.~A.} \bibnamefont{Telnov}},
  \bibnamefont{and} \bibinfo{author}{\bibfnamefont{G.}~\bibnamefont{Plunien}},
  \bibinfo{journal}{The European Physical Journal D}
  \textbf{\bibinfo{volume}{74}}, \bibinfo{pages}{188} (\bibinfo{year}{2020}).

\bibitem[{\citenamefont{Vembe et~al.}(2024)\citenamefont{Vembe, Johnsen, and
  F\o{}rre}}]{Vembe2024}
\bibinfo{author}{\bibfnamefont{J.~E.} \bibnamefont{Vembe}},
  \bibinfo{author}{\bibfnamefont{E.~A.~B.} \bibnamefont{Johnsen}},
  \bibnamefont{and} \bibinfo{author}{\bibfnamefont{M.}~\bibnamefont{F\o{}rre}},
  \bibinfo{journal}{Phys. Rev. A} \textbf{\bibinfo{volume}{109}},
  \bibinfo{pages}{013107} (\bibinfo{year}{2024}).

\bibitem[{\citenamefont{Fillion-Gourdeau
  et~al.}(2017)\citenamefont{Fillion-Gourdeau, MacLean, and
  Laflamme}}]{FillionGourdeau17}
\bibinfo{author}{\bibfnamefont{F.}~\bibnamefont{Fillion-Gourdeau}},
  \bibinfo{author}{\bibfnamefont{S.}~\bibnamefont{MacLean}}, \bibnamefont{and}
  \bibinfo{author}{\bibfnamefont{R.}~\bibnamefont{Laflamme}},
  \bibinfo{journal}{Phys. Rev. A} \textbf{\bibinfo{volume}{95}},
  \bibinfo{pages}{042343} (\bibinfo{year}{2017}).

\bibitem[{\citenamefont{Blanes et~al.}(2009)\citenamefont{Blanes, Casas, Oteo,
  and Ros}}]{Blanes2009}
\bibinfo{author}{\bibfnamefont{S.}~\bibnamefont{Blanes}},
  \bibinfo{author}{\bibfnamefont{F.}~\bibnamefont{Casas}},
  \bibinfo{author}{\bibfnamefont{J.}~\bibnamefont{Oteo}}, \bibnamefont{and}
  \bibinfo{author}{\bibfnamefont{J.}~\bibnamefont{Ros}},
  \bibinfo{journal}{Physics Reports} \textbf{\bibinfo{volume}{470}},
  \bibinfo{pages}{151 } (\bibinfo{year}{2009}).

\bibitem[{\citenamefont{Selst\o{} et~al.}(2009)\citenamefont{Selst\o{},
  Lindroth, and Bengtsson}}]{Selsto2009}
\bibinfo{author}{\bibfnamefont{S.}~\bibnamefont{Selst\o{}}},
  \bibinfo{author}{\bibfnamefont{E.}~\bibnamefont{Lindroth}}, \bibnamefont{and}
  \bibinfo{author}{\bibfnamefont{J.}~\bibnamefont{Bengtsson}},
  \bibinfo{journal}{Phys. Rev. A} \textbf{\bibinfo{volume}{79}},
  \bibinfo{pages}{043418} (\bibinfo{year}{2009}).

\bibitem[{\citenamefont{Telnov and Chu}(2020)}]{Telnov2020}
\bibinfo{author}{\bibfnamefont{D.~A.} \bibnamefont{Telnov}} \bibnamefont{and}
  \bibinfo{author}{\bibfnamefont{S.-I.} \bibnamefont{Chu}},
  \bibinfo{journal}{Phys. Rev. A} \textbf{\bibinfo{volume}{102}},
  \bibinfo{pages}{063109} (\bibinfo{year}{2020}).

\bibitem[{\citenamefont{V\'azquez~de Aldana
  et~al.}(2001)\citenamefont{V\'azquez~de Aldana, Kylstra, Roso, Knight, Patel,
  and Worthington}}]{Aldana2001}
\bibinfo{author}{\bibfnamefont{J.~R.} \bibnamefont{V\'azquez~de Aldana}},
  \bibinfo{author}{\bibfnamefont{N.~J.} \bibnamefont{Kylstra}},
  \bibinfo{author}{\bibfnamefont{L.}~\bibnamefont{Roso}},
  \bibinfo{author}{\bibfnamefont{P.~L.} \bibnamefont{Knight}},
  \bibinfo{author}{\bibfnamefont{A.}~\bibnamefont{Patel}}, \bibnamefont{and}
  \bibinfo{author}{\bibfnamefont{R.~A.} \bibnamefont{Worthington}},
  \bibinfo{journal}{Phys. Rev. A} \textbf{\bibinfo{volume}{64}},
  \bibinfo{pages}{013411} (\bibinfo{year}{2001}).

\bibitem[{\citenamefont{F\o{}rre and Simonsen}(2016)}]{Forre16}
\bibinfo{author}{\bibfnamefont{M.}~\bibnamefont{F\o{}rre}} \bibnamefont{and}
  \bibinfo{author}{\bibfnamefont{A.~S.} \bibnamefont{Simonsen}},
  \bibinfo{journal}{Phys. Rev. A} \textbf{\bibinfo{volume}{93}},
  \bibinfo{pages}{013423} (\bibinfo{year}{2016}).

\bibitem[{\citenamefont{Simonsen and F\o{}rre}(2016)}]{Simonsen2016}
\bibinfo{author}{\bibfnamefont{A.~S.} \bibnamefont{Simonsen}} \bibnamefont{and}
  \bibinfo{author}{\bibfnamefont{M.}~\bibnamefont{F\o{}rre}},
  \bibinfo{journal}{Phys. Rev. A} \textbf{\bibinfo{volume}{93}},
  \bibinfo{pages}{063425} (\bibinfo{year}{2016}).

\bibitem[{\citenamefont{F\o{}rre and Selst\o{}}(2020)}]{Forre2020}
\bibinfo{author}{\bibfnamefont{M.}~\bibnamefont{F\o{}rre}} \bibnamefont{and}
  \bibinfo{author}{\bibfnamefont{S.}~\bibnamefont{Selst\o{}}},
  \bibinfo{journal}{Phys. Rev. A} \textbf{\bibinfo{volume}{101}},
  \bibinfo{pages}{063416} (\bibinfo{year}{2020}).

\bibitem[{\citenamefont{Suster et~al.}(2023)\citenamefont{Suster, Derlikiewicz,
  Krajewska, V\'elez, and Kami\ifmmode~\acute{n}\else
  \'{n}\fi{}ski}}]{Suster2023}
\bibinfo{author}{\bibfnamefont{M.~C.} \bibnamefont{Suster}},
  \bibinfo{author}{\bibfnamefont{J.}~\bibnamefont{Derlikiewicz}},
  \bibinfo{author}{\bibfnamefont{K.}~\bibnamefont{Krajewska}},
  \bibinfo{author}{\bibfnamefont{F.~C.} \bibnamefont{V\'elez}},
  \bibnamefont{and} \bibinfo{author}{\bibfnamefont{J.~Z.}
  \bibnamefont{Kami\ifmmode~\acute{n}\else \'{n}\fi{}ski}},
  \bibinfo{journal}{Phys. Rev. A} \textbf{\bibinfo{volume}{107}},
  \bibinfo{pages}{053112} (\bibinfo{year}{2023}).

\bibitem[{\citenamefont{Suster et~al.}(2024)\citenamefont{Suster, Derlikiewicz,
  Kami\'{n}ski, and Krajewska}}]{Suster2024}
\bibinfo{author}{\bibfnamefont{M.~C.} \bibnamefont{Suster}},
  \bibinfo{author}{\bibfnamefont{J.}~\bibnamefont{Derlikiewicz}},
  \bibinfo{author}{\bibfnamefont{J.~Z.} \bibnamefont{Kami\'{n}ski}},
  \bibnamefont{and}
  \bibinfo{author}{\bibfnamefont{K.}~\bibnamefont{Krajewska}},
  \bibinfo{journal}{Opt. Express} \textbf{\bibinfo{volume}{32}},
  \bibinfo{pages}{6085} (\bibinfo{year}{2024}).

\bibitem[{\citenamefont{Lindblom et~al.}(2020)\citenamefont{Lindblom, F\o{}rre,
  Lindroth, and Selst\o{}}}]{Lindblom2020}
\bibinfo{author}{\bibfnamefont{T.~K.} \bibnamefont{Lindblom}},
  \bibinfo{author}{\bibfnamefont{M.}~\bibnamefont{F\o{}rre}},
  \bibinfo{author}{\bibfnamefont{E.}~\bibnamefont{Lindroth}}, \bibnamefont{and}
  \bibinfo{author}{\bibfnamefont{S.}~\bibnamefont{Selst\o{}}},
  \bibinfo{journal}{Phys. Rev. A} \textbf{\bibinfo{volume}{102}},
  \bibinfo{pages}{063108} (\bibinfo{year}{2020}).

\bibitem[{\citenamefont{F\o{}rre}(2019)}]{Forre2019}
\bibinfo{author}{\bibfnamefont{M.}~\bibnamefont{F\o{}rre}},
  \bibinfo{journal}{Phys. Rev. A} \textbf{\bibinfo{volume}{99}},
  \bibinfo{pages}{053410} (\bibinfo{year}{2019}).

\bibitem[{\citenamefont{Kjellsson~Lindblom and Selst\o{}}(2021)}]{Lindblom2021}
\bibinfo{author}{\bibfnamefont{T.}~\bibnamefont{Kjellsson~Lindblom}}
  \bibnamefont{and}
  \bibinfo{author}{\bibfnamefont{S.}~\bibnamefont{Selst\o{}}},
  \bibinfo{journal}{Phys. Rev. A} \textbf{\bibinfo{volume}{104}},
  \bibinfo{pages}{043102} (\bibinfo{year}{2021}).

\bibitem[{\citenamefont{Sonnleitner and Barnett}(2018)}]{Sonnleitner18}
\bibinfo{author}{\bibfnamefont{M.}~\bibnamefont{Sonnleitner}} \bibnamefont{and}
  \bibinfo{author}{\bibfnamefont{S.~M.} \bibnamefont{Barnett}},
  \bibinfo{journal}{Phys. Rev. A} \textbf{\bibinfo{volume}{98}},
  \bibinfo{pages}{042106} (\bibinfo{year}{2018}).

\bibitem[{\citenamefont{Foldy and Wouthuysen}(1950)}]{Foldy50}
\bibinfo{author}{\bibfnamefont{L.~L.} \bibnamefont{Foldy}} \bibnamefont{and}
  \bibinfo{author}{\bibfnamefont{S.~A.} \bibnamefont{Wouthuysen}},
  \bibinfo{journal}{Phys. Rev.} \textbf{\bibinfo{volume}{78}},
  \bibinfo{pages}{29} (\bibinfo{year}{1950}).

\bibitem[{\citenamefont{Madsen and Lambropoulos}(1999)}]{Lambropoulos1999}
\bibinfo{author}{\bibfnamefont{L.~B.} \bibnamefont{Madsen}} \bibnamefont{and}
  \bibinfo{author}{\bibfnamefont{P.}~\bibnamefont{Lambropoulos}},
  \bibinfo{journal}{Phys. Rev. A} \textbf{\bibinfo{volume}{59}},
  \bibinfo{pages}{4574} (\bibinfo{year}{1999}).

\bibitem[{\citenamefont{deBoor}(1978)}]{deboor}
\bibinfo{author}{\bibfnamefont{C.}~\bibnamefont{deBoor}},
  \emph{\bibinfo{title}{A Practical Guide to Splines}}
  (\bibinfo{publisher}{Springer-Verlag}, \bibinfo{address}{New York},
  \bibinfo{year}{1978}).

\bibitem[{\citenamefont{Vanne and Saenz}(2012)}]{Vanne2012}
\bibinfo{author}{\bibfnamefont{Y.~V.} \bibnamefont{Vanne}} \bibnamefont{and}
  \bibinfo{author}{\bibfnamefont{A.}~\bibnamefont{Saenz}},
  \bibinfo{journal}{Phys. Rev. A} \textbf{\bibinfo{volume}{85}},
  \bibinfo{pages}{033411} (\bibinfo{year}{2012}).

\bibitem[{\citenamefont{Fischer and Zatsarinny}(2009)}]{Fischer2009}
\bibinfo{author}{\bibfnamefont{C.~F.} \bibnamefont{Fischer}} \bibnamefont{and}
  \bibinfo{author}{\bibfnamefont{O.}~\bibnamefont{Zatsarinny}},
  \bibinfo{journal}{Comput. Phys. Commun.} \textbf{\bibinfo{volume}{180}},
  \bibinfo{pages}{879 } (\bibinfo{year}{2009}).

\bibitem[{\citenamefont{Saad}(1992)}]{Saad1992}
\bibinfo{author}{\bibfnamefont{Y.}~\bibnamefont{Saad}}, \bibinfo{journal}{SIAM
  J. Numer. Anal.} \textbf{\bibinfo{volume}{29}}, \bibinfo{pages}{209}
  (\bibinfo{year}{1992}).

\end{thebibliography}
\end{document}